\documentclass[onecolumn,referee,traditabstract]{aa}

\usepackage{natbib,soul}
\usepackage[english]{babel}
\usepackage{algorithmic}
\usepackage{algorithm}
\usepackage{theorem}
\usepackage{amssymb}
\usepackage{amsmath}
\usepackage{mathtools}
\usepackage{graphics}
\usepackage{subfigure}
\usepackage{fullpage}
\usepackage{url}
\usepackage{hyperref,ulem,enumerate}
\usepackage{epsfig}
\newcommand{\bibfont}{\tiny}

\def\n{\mathbf{n}}

\graphicspath{./figures/} 

\usepackage{amstext,amssymb,amsbsy,amsmath,amscd,mathrsfs,dsfont}
\usepackage{graphics,graphicx,epsfig,subfigure,enumerate,rotating,theorem,algorithm,algorithmic,array,longtable}

\newcommand{\be}{\begin{eqnarray}}
\newcommand{\ee}{\end{eqnarray}}

\newcommand{\norm}[1]{\left\|#1\right\|}

\newcommand{\abs}[1]{\left|#1\right|}

\newcommand{\beqn}{\begin{eqnarray}}
\newcommand{\eeqn}{\end{eqnarray}}

\renewcommand{\leq}{\leqslant}

\newcommand{\argmin}[1]{\underset{#1}{\mathrm{argmin}}\;}

\begin{document}

\title{Sparsity and the Bayesian Perspective}

\author{J.-L. Starck  \inst{1} and D.L. Donoho  \inst{2} and  M.J. Fadili \inst{3} and A. Rassat \inst{3,1}}

\institute{
\inst{1} Laboratoire AIM, UMR CEA-CNRS-Paris 7, Irfu, Service d'Astrophysique, CEA Saclay, F-91191 GIF-SUR-YVETTE Cedex, France  \\
\inst{2}  Department of Statistics, Stanford University, Stanford CA, 94305, USA.  \\
\inst{3} GREYC CNRS-ENSICAEN-Universit\'e de Caen, 6, Bd du Mar\'echal Juin, 14050 Caen Cedex, France\\
\inst{3} Laboratoire d'Astrophysique, Ecole Polytechnique F\'ed\'erale de Lausanne (EPFL), Observatoire de Sauverny, CH-1290, Versoix, Switzerland}
 
\date\today

\abstract{Sparsity has been recently introduced in cosmology for weak-lensing and CMB data analysis for different applications such as denoising, component separation or inpainting (i.e. filling the missing data or the mask).
Although it gives very nice numerical results, CMB sparse inpainting has been severely criticized by top researchers in cosmology, based on arguments derived from a Bayesian perspective.
Trying to understand their point of view, we realize that interpreting a regularization penalty term as a prior in a Bayesian framework can lead to erroneous conclusions.
This paper is by no means against the Bayesian approach, which has proven to be very useful for many applications, 
but warns about a Bayesian-only interpretation in data analysis, which can be misleading in some cases.}
 

\maketitle 

 
\section{Introduction}
Bayesian methodology is extremely popular in cosmology. It proposes a very elegant framework to deal with uncertainties and to use our knowledge, under the form of priors,
in order to solve a given inverse problem \citep{Hobson,Mielczarek}. The huge success of Bayesian methods in cosmology is well illustrated in \citet{trotta2008bayes}
with a figure on the number of papers with the word "Bayesian" in the title as a function of the publication year. Hence Bayesian techniques have been  used for many applications such as 
model selection \citep{kilbinger2010bayesian,Trotta:2012fk}, primordial power spectrum analysis  \citep{kawasaki2010probing}, galactic surveys design  \citep{Watkinson}, or
 cosmological parameters estimations \citep{March}. 
 Bayesian methods are now commonly used at almost every step in the CMB pipeline experiments, for point sources 
 removal \citep{Argueso,Carvalho}, noise level estimation \citep{Wehus},  component separation \citep{Dickinson}, cosmological parameter estimation \citep{Efstathiou}, 
non-Gaussianity studies \citep{Elsner,Feeney}, or inpainting  \citep{bucher2012,kim2012}.

Sparsity has recently  been proposed for CMB data analysis  for component separation \citep{bobin2012} and inpainting \citep{inpainting:abrial06,starck:abrial08,starck2012}.
The sparsity-based inpainting approach has been successfully used for two different CMB studies, the CMB weak-lensing on  Planck simulated data \citep{perotto10,stephane2012},
and the analysis of the integrated Sachs-Wolfe effect (ISW) on WMAP data  \citep{starck:dupe2011,starck:rassat2012}. In both cases, the authors showed from Monte-Carlo simulations that the statistics derived from the inpainted maps can be trusted at a high confidence level, and that sparsity-based inpainting can indeed provide an easy and effective solution to the large Galactic mask problem. 

However, even if these simulations have shown that the sparsity-based inpainting does not destroy CMB  weak-lensing, ISW signals or some large-scale anomalies in the CMB, 
the CMB community is very reluctant to use this concept.    

This has lead to very animated discussions in conferences.  During these discussions it arose that cosmologists often resorted to a Bayesian interpretation of sparsity. 
Hence, they {summarize} the sparse regularization to the Maximum A Posteriori (MAP) estimator  assuming the solution follows a Laplacian distribution.
From this perspective, several strong arguments against the use of sparsity for CMB analysis  were raised:
\begin{enumerate}
\item Sparsity consists in assuming an anisotropic and a  non-Gaussian prior, which {is unsuitable} for the CMB, which is Gaussian and isotropic.
\item Sparsity violates rotational invariance.
\item The $\ell_1$ norm that is used for sparse inpainting arose purely out of expediency because under certain circumstances it reproduces the results of the $\ell_0$ pseudo-norm   
(which arises naturally in the context of strict, as opposed to weak, sparsity) without necessitating combinatorial optimization. 
\item There is no mathematical proof that sparse regularization preserves/recovers the original statistics.
\end{enumerate}

The above arguments result from a Bayesian point of view of the sparsity concept. 
In this paper we explain in detail how the above arguments are actually not rigorously valid and how sparsity is not in contradiction with a Bayesian interpretation. 


\section{Sparse Regularization of Inverse Problems}

\subsection{Inverse Problem Regularization}
Many data processing problems can be formalized as a linear inverse problem, 
\begin{equation}
\label{eq_pbinv}
y = A x + \varepsilon ~,
\end{equation}
where $y \in \mathbb{F}^m$ is a vector of noisy measurements (real with $\mathbb{F}=\mathbb{R}$, or complex $\mathbb{F}=\mathbb{C}$), $\varepsilon$ is an $m$-dimensional vector of additive
noise, $x$ is the perfect $n$-dimensional unobserved vector of interest, 
and $A: \mathbb{F}^n \to \mathbb{F}^m$ is a linear operator. 
For example, the inpainting problem corresponds to the case where we want to recover some missing data, 
 in which case $A$ is a binary matrix with less rows than columns ($m < n$), and it contains only one value equal to $1$ per row, and all other values equal to $0$. 
 
Finding $x$ knowing the data $y$ and $A$ is a linear inverse problem. 
When it does not have a unique and stable solution, it is {\it ill-posed}, and a regularization is necessary to reduce the space of candidate solutions. 
 
A very popular regularization in astronomy is the well-known Bayesian Maximum Entropy Method (MEM), which is based on the principle that we want to select the simplest solution which fits the data. Sparsity has recently emerged as very powerful approach for regularization \citep{starck:book10}.

\subsection{Strict and Weak Sparsity}
A signal $x$ considered as a vector in $\mathbb{F}^n$ is said to be sparse if
most of its entries are equal to zero. If $k$ numbers of the $n$ samples are equal to zero, where $k <  n$, then the signal is said to be $k$-sparse. 
Generally signals are not sparse in direct space, but can be sparsified by transforming
them to another domain. Think for instance of a purely sinusoidal signal which is $1$-sparse
in the Fourier domain, while it is clearly dense in the original
one. In the so-called sparsity synthesis model, a signal can be represented as the linear expansion 
\begin{equation}
x =\Phi\alpha=\sum_{i=1}^{t}\phi_{i}\alpha_i ~,
\end{equation}
where $\alpha$ are the synthesis coefficient of $x$, $\Phi=(\phi_{1},\ldots,\phi_{t})$ is the dictionary whose columns
are $t$ elementary waveforms $\phi_{i}$ also called atoms.
 In the language of linear algebra, the dictionary $\Phi$ is a $b\times t$ matrix whose columns are the atoms normalized, supposed here to be normalized to a unit $\ell_{2}$-norm, i.e.  $\forall i\in[1,t],\left\Vert \phi_{i}\right\Vert _2^2=\sum_{n=1}^{n}\left|\phi_{i}[n]\right|^{2}=1$.

A signal can be decomposed in many dictionaries, but the best one is that with the 
sparsest (most economical) representation of the signal. 
In practice, it is convenient to use dictionaries with fast implicit transform (such as Fourier transform, wavelet transform, etc.) which allow us to directly obtain the coefficients and reconstruct the signal from these coefficients using fast algorithms running in linear or almost linear time (unlike matrix-vector multiplications).
The Fourier, wavelet and discrete cosine transforms are among the most popular dictionaries.

Most natural signals however are not exactly sparse but rather concentrated near a small set.
Such signals are termed {\it compressible} or {\it weakly sparse}
in the sense that the sorted magnitudes $  \abs{ \alpha_{(i)}  }$, i.e.  $  \abs{ \alpha_{(1)}  } >  \abs{ \alpha_{(2)}}, ...,   > \abs{ \alpha_{(t)}}$,
 of the sequence of coefficients  $\alpha$ decay quickly according 
 to a power law, i.e.  $\abs{ \alpha_{(i)}}  \leq C i^{-1/r} ~, i=1,\ldots,t $,  where $C$ 
 is a constant.The larger $r$ is, the faster the coefficients amplitudes decay, and the more compressible the signal is. In turn, the non-linear $\ell_2$ approximation error of $\alpha$ (and $x$) from its $M$ largest entries in magnitude decrease also quickly. One can think for instance of the wavelet coefficients of a smooth signal away from isotropic singularities, or the curvelet coefficients of a piecewise regular image away from smooth contours. A comprehensive account on sparsity of signals and images can be found in \citep{starck:book10}.

\subsection{Sparse  Regularization for Inverse Problems}

In the following, for a vector $z$ we denote $\| z \|^p_p=\sum_i |z_i|^p$ for $p \geq 0$. In particular, for $p \geq 1$, this is the $p$-th power of the $\ell_p$ norm, and for $p=0$, we get the $\ell_0$ pseudo-norm which counts the number of non-zero entries in $z$. The $\ell_0$ regularized problem amounts to minimizing

\begin{equation}
\tilde \alpha \in \argmin{\alpha}     \| y - A  \Phi \alpha  \|_{2}^2 + \lambda    \| \alpha \|_0 ,
\label{minimisationl0}
\end{equation}
where $\lambda$ is a regularization parameter.  A solution $\tilde x$ is reconstructed as  $\tilde x =  \Phi \tilde \alpha$. Clearly, the goal of \eqref{minimisationl0} is to minimize the number of non-zero coefficients describing the sought after signal while ensuring that the forward model is faithful to the observations.


Solving \eqref{minimisationl0} is however known to be NP-hard. The $\ell_1$ norm has then been proposed as a tight convex relaxation of  \eqref{minimisationl0} leading to the minimization problem
\begin{equation}
\label{eq_l1_inv}
\tilde \alpha \in \argmin{\alpha}     \| y - A  \Phi \alpha  \|_{2}^2 + \lambda    \| \alpha \|_1 ,
\end{equation}
where $\lambda$ is again a regularization parameter different from that of \eqref{minimisationl0}. There has been a tremendous amount of work where researchers spanning a wide range of disciplines have studied the structural properties of minimizers of \eqref{eq_l1_inv} and its equivalence with \eqref{minimisationl0}. Not only problem \eqref{eq_l1_inv} is computationally appealing and can be efficiently solved, but it has been proved that under appropriate circumstances, \eqref{eq_l1_inv}  produces exactly the same solutions as \eqref{minimisationl0}, see e.g. \citep{DonohoMost} and the  overview in the monograph \citep{starck:book10}.


\section{Sparsity Prior and  Bayesian Prior}
\subsection{Bayesian framework}
In the Bayesian framework, a prior is imposed on the object of interest through a probability distribution. For instance, assume that the coefficients $\alpha$ are i.i.d. Laplacian with the scale parameter $\tau$, i.e. the density $P_\alpha(\alpha ) \propto e^{ - \tau\norm{\alpha}_1 }$, and the noise $\varepsilon$ is zero-mean white Gaussian with variance $\sigma^2$, i.e. the conditional density $P_{Y|\alpha}(y) =  (2\pi \sigma^2)^{-m/2}e^{ - \| y - A  \Phi \alpha  \|_2^2/(2\sigma^2)}$. By traditional Bayesian arguments, the Maximum {\it a posteriori} (MAP) estimator is obtained 
 by maximizing the conditional posterior density $P_{\alpha|Y}(\alpha)  \propto  P_{Y | \alpha}(y) P_{\alpha}(\alpha)$, or equivalently by minimizing its anti-log version
\begin{equation}
 \min_{\alpha}  \frac{1}{2\sigma^2}\norm{ y - A  \Phi \alpha  }_2^2  +  \tau \norm{\alpha}_1.
\end{equation}
This is exactly \eqref{eq_l1_inv} by identifying $\lambda=2\sigma^2\tau$.
This resemblance has led Bayesian cosmologists to raise the four criticisms mentioned in the introduction. But as we will discuss shortly, their central argument is not used in the right sense which can yield misleading conclusions.

\subsection{Should $\ell_1$ regularization be the MAP ?}
In Bayesian cosmology, the following shortcut is often made: if a prior {is at the basis of an} algorithm, then to use this algorithm, the resulting coefficients must be distributed according to this prior. But it is a false logical chain in general and high-dimensional phenomena completely invalidate it.

For instance, Bayesian cosmologists claim that $\ell_1$ regularization is \textit{equivalent} to assuming that the solution is Laplacian and not Gaussian, which {would be unsuitable for} the case of CMB analysis. This argument however assumes that a MAP estimate follows the distribution of the prior. But it is now well-established that MAP solutions substantially deviate from the prior model, and that the {disparity} between the prior and the effective distribution obtained from the true MAP estimate is a permanent contradiction in Bayesian MAP estimation \citep{nikolova2007}. Even the supposedly correct $\ell_2$ prior would yield an estimate (Wiener which coincides with the MAP and posterior conditional mean), whose covariance is not that of the prior.

In addition, rigorously speaking, this MAP interpretation of $\ell_1$ regularization is not the only possible one. More precisely, it was shown in \citet{gribonval2012,gribonval2011should,baraniuk10} that solving
a penalized least squares regression problem with penalty $\psi(\alpha)$ (e.g. the $\ell_1$ norm) should not necessarily be interpreted as assuming a Gibbsian prior $C \exp(-\psi(\alpha))$ and using the MAP estimator. In particular, for any prior $P_{\alpha}$, the conditional mean can be equally interpreted as a MAP with some prior $C \exp(-\psi(\alpha))$. Conversely, for certain penalties $\psi(\alpha)$, the solution of the penalized least squares problem is indeed the conditional posterior mean, with a certain prior $P_\alpha(\alpha)$ which is in general different from $C \exp(-\psi(\alpha))$.

In summary, the MAP interpretation of such penalized least-squares regression can be misleading, and using a MAP estimation, the solution does not necessary follow the prior distribution, and an {\it incorrect} prior does not necessarily lead to a wrong solution. What we are claiming here are facts that were stated and proved as rigorous theorems in the literature.
  
 
\subsection{Compressed sensing: the Bayesian interpretation {inadequacy} }
A beautiful example to illustrate this is the compressed {sensing} scenario \citep{donoho:cs,CandesTao04}, 
which tells us that a $k$-sparse, or compressible, $n$-dimensional signal $x$ can
be recovered either exactly or to a good approximation from much less random measurements $m$ than the ambient dimension $n$, 
if $m$ is sufficiently larger than the intrinsic dimension of $x$.
Clearly, the underdetermined linear problem, $y = Ax$, where $A$ is drawn from an appropriate random ensemble, with less equations than unknown, {\it can} be solved exactly or approximately, if the underlying object $x$ is sparse or compressible. This can be furthermore achieved by solving a computationally tractable $\ell_1$-regularized convex optimization program.
 
If the underlying signal is exactly sparse, in a Bayesian framework, this would be a completely absurd way to solve the problem, since the Laplacian prior is very different from the actual properties of the original signal (i.e. $k$ coefficients different from zero). In particular, what compressed sensing shows, is that we can have prior A be completely true, but utterly impossible to use for computation time or any other reason, 
and can use prior B instead, and get the correct results! Therefore, from a  Bayesian point of view, it is rather difficult to understand not only that the $\ell_1$ norm is adequate, but that it also leads to the exact solution.

\subsection{The prior misunderstanding}
Considering the chosen dictionary $\Phi$ as the spherical harmonic transform, then the coefficients $\alpha$ are now $\alpha = \left\{  a_{l,m} \right\}_{l=0,\ldots,l_{\max}, m=-l,\ldots,l}$. The standard CMB theory assumes that the 
$a_{l,m}$'s are the realizations of a collection of heteroscedastic complex zero-mean Gaussian random variables with variances $C_l / 2$, where $C_l$ is the true power spectrum.
Using a $\ell_1$ norm is then interpreted in the 
Bayesian framework as having a Laplacian prior on each $a_{l,m}$ 
which is in contradiction with the underlying CMB theory.

\begin{figure*}[htb]
\vbox{
\hbox{
\includegraphics [scale=0.5]{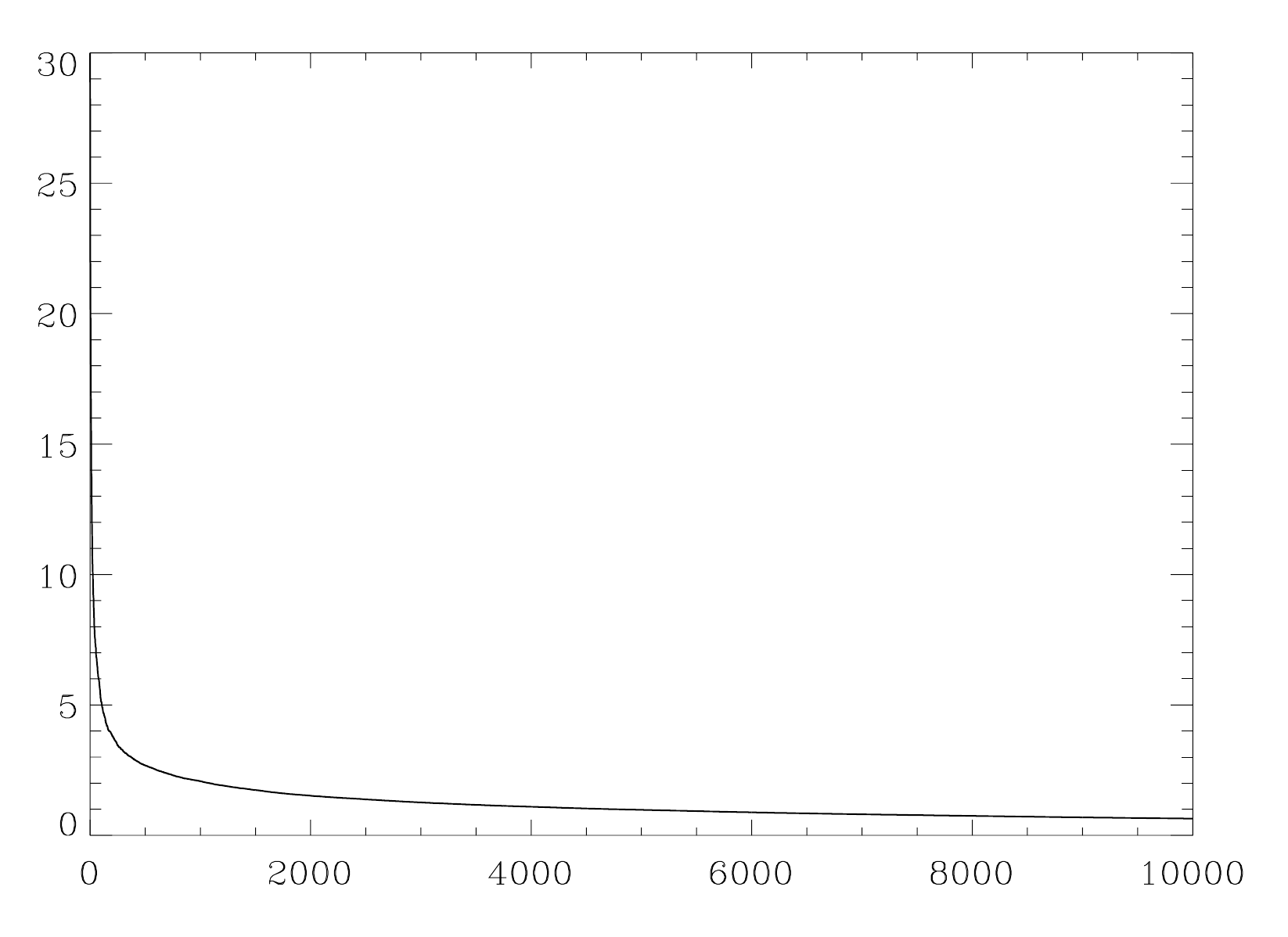}
\includegraphics [scale=0.5]{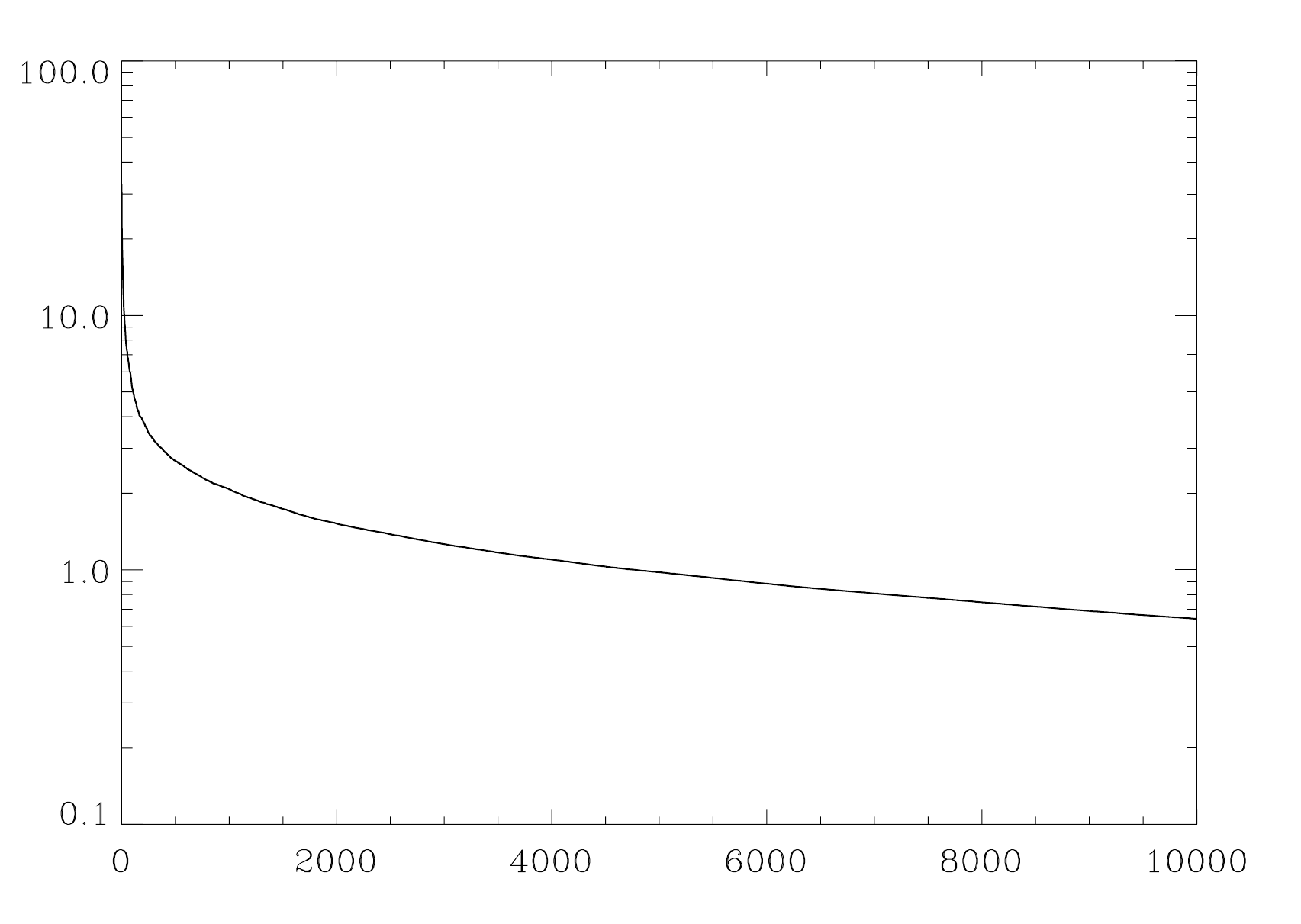}
}}
\caption{Left: amplitude (absolute value) of the spherical harmonic coefficients versus their index, when the coefficients are sorted from the largest amplitude to the smallest. 
Right, same plot with a y-axis in log.}
\label{fig_decay_cmb_wt}
\end{figure*}

However, as argued above, from the regularization point of view, the $\ell_1$ norm merely promotes a solution $x$ such that its spherical harmonic coefficients are (nearly) sparse.
 There is no assumption at all on the properties related to a specific $a_{l,m}$. In fact, there is no randomness put in here and the $a_{l,m}$ values does not even 
 have to be interpreted as a realization of a random variable.
 Therefore, whatever the underlying distribution for a given $a_{l,m}$ (if its exists), it needs not be interpreted as Laplacian under the $\ell_1$ regularization. The CMB can be Gaussian or not, isotropic or not, and still there will be no contradiction with the principle of using the $\ell_1$ norm to solve the reconstruction problem (e.g. inpainting).  
What is important is that the sorted absolute values of the CMB spherical harmonics coefficients presents a fast decay.
This is easily verified using a CMB map, data or simulation. 
This is well illustrated by the  Fig.~\ref{fig_decay_cmb_wt} which shows this decay for the Nine-Year WMAP data set. 
As we can see, the compressibility assumption is completely valid.
 
\section{Discussion of the Bayesian criticisms}
\begin{enumerate}[$\bullet$]
\item{\it Sparsity consists in assuming an anisotropy and a non Gaussian prior, {which is unsuitable} for the CMB, which is Gaussian and isotropic.}
We have explained in the previous section that, not only this MAP interpretation of $\ell_1$ regularization is misleading, but also that there is no assumption at all on the underlying stochastic process. The CMB sparsity-based recovery is a purely data-driven regularization approach:  
the sorted absolute values of the spherical harmonic coefficients presents a fast decay, as seen on real data or simulations, and this motivates the sparse regularizations.

There is no assumption  that the CMB is Gaussian or isotropic, but there is also no assumption that it is not Gaussian or anisotropic. 
In this sense, using the $\ell_1$-regularized inpainting to test if the CMB is indeed Gaussian 
and isotropic may be better than other methods, including Wiener filtering, which in the Bayesian framework assumes Gaussianity and isotropy. Furthermore, the Wiener estimator will also require to know the underlying power spectrum (i.e. the theoretical $C_{l}$) which is an even stronger prior.

\item {\it Sparsity violates rotational invariance.}
The criticicism here is that linear combinations of independent exponentials are not independent exponentials; hence isotropy is necessarily violated, unless the $a_{lm}$ are Gaussian.
But again, our arguments for $\ell_1$ regularization are borrowed from approximation theory and harmonic analysis, and this does not contradict that the $a_{lm}$ coefficients can be realizations of a sequence of heteroscedastic Gaussian variables.
The $\ell_1$ norm regularization will be adequate if the sorted coefficients amplitude follow a fast decay, which is always verified with CMB data.
Indeed, as we already mentioned, a set of parameters $x_i$, were each $x_i$ is a realization of a Gaussian process of mean $0$ and variance $V_i$, can present a fast decay
when we plot the sorted absolute values of $x_i$. In case of CMB spherical harmonic coefficients, it is straightforward to verify this.

\item {\it The $\ell_1$ norm that is used for sparse inpainting arose purely out of expediency because under certain circumstances it reproduces the results of the $\ell_0$ pseudo-norm.}
First, we would like to correct a possible misunderstanding: $\ell_1$ regularization can provably recover both strictly and weakly sparse signals, while being stable to noise. In the strictly sparse case, $\ell_1$ norm minimization can recover the right solution although the prior is not correct from a Bayesian point of view. In the compressible case, the recovery is up to a good approximation, as good as an oracle that would give the best $M$-term approximation of the unknown signal (i.e. $\ell_0$-solution). What is criticized as a "expedient" prior is basically at the heart of regularization theory, for instance here $\ell_1$, while providing strong theoretical recovery guarantees under appropriate conditions. A closer look at the literature of inverse problems shows that these guarantees are possible beyond the compressed sensing scenario.

\item {\it There is no mathematical proof that sparse regularization preserves/recovers the original statistics.}
This is true, but this arguments is not specific to $\ell_1$ regularized inpainting. Even worse, as detailed and argued in the previous section, not only the posterior distribution generally deviates from the prior, but even if one uses the MAP with the correct Gaussian prior, the MAP estimate (Wiener) will not have the covariance of the prior. 

Another point to mention is that the CMB is never a pure Gaussian random field, even if the standard $\Lambda$CDM is truly valid. Indeed, we know that the CMB is at least contaminated by {non-Gaussian} secondary anisotropies such as weak-lensing effects or kinetic SZ. Therefore the original CMB statistics are likely better preserved by an inpainting method that does not assume Gaussianity (but nonetheless allows it), rather than by a method which has an explicit Gaussian assumption. Moreover, if the CMB is not Gaussian, then one can clearly anticipate that the Wiener estimate does not preserve the original statistics. 

Finally, it appears unfair to criticize sparsity-regularised inverse problems on the mathematical side. Indeed, a quick look at the literature shows the vitality of the mathematical community (pure and applied) and the large amount of theoretical guarantees (deterministic and frequentist statistical settings) that have been devoted to $\ell_1$ regularization. In particular, theoretical guarantees of $\ell_1$-based inpainting can be found in \citep{king12} on the Cartesian grid, and of sparse recovery on the sphere in \citep{rauhut10}.


\end{enumerate}


\section{Conclusions}
In Bayesian cosmology, the following shortcut is often made: 
if a prior {is at the basis of an} algorithm, then to use this algorithm, the resulting coefficients must be distributed according to this prior.
But it is a false logical chain, and as discussed above, high-dimensional phenomena completely invalidate it. 
In particular, what compressed sensing shows, is that we can have prior A be completely true, but utterly impossible to use for computation time or any other reason, 
and can use prior B instead, and get the correct results! 
We don't claim in this paper that the sparse inpainting is the best solution for inpainting, but we showed that arguments raised against it are incorrect, and that if
Bayesian methodology offers a very elegant framework extremely useful for many applications, 
we should be careful not to be monolithic in the way we address a problem. 
In practice, it may be useful to use several inpainting methods to better understand the CMB statistics, and it is clear that sparsity based inpainting does not require to make any assumption about the Gaussianity nor the isotropy, and does not need to have a theoretical $C_\ell$ as an input.


\section*{Acknowledgment}
The authors thank Benjamin Wandelt, Mike Hobson, Jason McEwen, Domenico Marinucci, Hiranya Peiris, Roberto Trotta, and 
Tom Loredo  for the useful and animated discussions. 
This work was supported by the French National Agency for Research (ANR -08-EMER-009-01),  
the European Research Council grant SparseAstro (ERC-228261), and the Swiss National Science Foundation (SNSF).

\bibliographystyle{aa}
\bibliography{BayesSparse}

\begin{thebibliography}{35}
\expandafter\ifx\csname natexlab\endcsname\relax\def\natexlab#1{#1}\fi

\bibitem[{{Abrial} {et~al.}(2007){Abrial}, {Moudden}, {Starck}, {Bobin},
  {Fadili}, {Afeyan}, \& {Nguyen}}]{inpainting:abrial06}
{Abrial}, P., {Moudden}, Y., {Starck}, J., {et~al.} 2007, Journal of Fourier
  Analysis and Applications, 13, 729--748

\bibitem[{Abrial {et~al.}(2008)Abrial, Moudden, Starck, Fadili, Delabrouille,
  \& Nguyen}]{starck:abrial08}
Abrial, P., Moudden, Y., Starck, J., {et~al.} 2008, Statistical Methodology, 5,
  289-298

\bibitem[{{Arg{\"u}eso} {et~al.}(2011){Arg{\"u}eso}, {Salerno}, {Herranz},
  {Sanz}, {Kuruo{\v g}lu}, \& {Kayabol}}]{Argueso}
{Arg{\"u}eso}, F., {Salerno}, E., {Herranz}, D., {et~al.} 2011, \mnras, 414,
  410

\bibitem[{Baraniuk {et~al.}(2010)Baraniuk, Cevher, \& Wakin}]{baraniuk10}
Baraniuk, R., Cevher, V., \& Wakin, M. 2010, Proceedings of the IEEE, 98, 959

\bibitem[{{Bobin} {et~al.}(2013){Bobin}, {Starck}, {Sureau}, \&
  {Basak}}]{bobin2012}
{Bobin}, J., {Starck}, J.-L., {Sureau}, F., \& {Basak}, S. 2013, \aap, 550, A73

\bibitem[{{Bucher} \& {Louis}(2012)}]{bucher2012}
{Bucher}, M. \& {Louis}, T. 2012, \mnras, 3271

\bibitem[{Cand{\`e}s \& Tao(2006)}]{CandesTao04}
Cand{\`e}s, E. \& Tao, T. 2006, {IEEE} Transactions on Information Theory, 52,
  5406--5425

\bibitem[{{Carvalho} {et~al.}(2011){Carvalho}, {Rocha}, {Hobson}, \&
  {Lasenby}}]{Carvalho}
{Carvalho}, P., {Rocha}, G., {Hobson}, M.~P., \& {Lasenby}, A. 2011, ArXiv
  e-prints

\bibitem[{{Dickinson} {et~al.}(2009){Dickinson}, {Eriksen}, {Banday}, {Jewell},
  {G{\'o}rski}, {Huey}, {Lawrence}, {O'Dwyer}, \& {Wandelt}}]{Dickinson}
{Dickinson}, C., {Eriksen}, H.~K., {Banday}, A.~J., {et~al.} 2009, \apj, 705,
  1607

\bibitem[{Donoho(2006{\natexlab{a}})}]{donoho:cs}
Donoho, D. 2006{\natexlab{a}}, IEEE Transactions on Information Theory, 52,
  1289--1306

\bibitem[{Donoho(2006{\natexlab{b}})}]{DonohoMost}
Donoho, D. 2006{\natexlab{b}}, Communications on Pure and Applied Mathematics,
  59, 907--934

\bibitem[{{Dup{\'e}} {et~al.}(2011){Dup{\'e}}, {Rassat}, {Starck}, \&
  {Fadili}}]{starck:dupe2011}
{Dup{\'e}}, F.-X., {Rassat}, A., {Starck}, J.-L., \& {Fadili}, M.~J. 2011,
  \aap, 534, A51

\bibitem[{{Efstathiou} {et~al.}(2010){Efstathiou}, {Ma}, \&
  {Hanson}}]{Efstathiou}
{Efstathiou}, G., {Ma}, Y.-Z., \& {Hanson}, D. 2010, \mnras, 407, 2530

\bibitem[{{Elsner} \& {Wandelt}(2010)}]{Elsner}
{Elsner}, F. \& {Wandelt}, B.~D. 2010, \apj, 724, 1262

\bibitem[{{Feeney} {et~al.}(2012){Feeney}, {Johnson}, {McEwen}, {Mortlock}, \&
  {Peiris}}]{Feeney}
{Feeney}, S.~M., {Johnson}, M.~C., {McEwen}, J.~D., {Mortlock}, D.~J., \&
  {Peiris}, H.~V. 2012, ArXiv e-prints

\bibitem[{Gribonval(2011)}]{gribonval2011should}
Gribonval, R. 2011, Signal Processing, IEEE Transactions on, 59, 2405

\bibitem[{{Gribonval} {et~al.}(2011){Gribonval}, {Cevher}, \&
  {Davies}}]{gribonval2012}
{Gribonval}, R., {Cevher}, V., \& {Davies}, M.~E. 2011, ArXiv 1102.1249

\bibitem[{{Hobson} {et~al.}(2010){Hobson}, {Jaffe}, {Liddle}, {Mukeherjee}, \&
  {Parkinson}}]{Hobson}
{Hobson}, M.~P., {Jaffe}, A.~H., {Liddle}, A.~R., {Mukeherjee}, P., \&
  {Parkinson}, D. 2010, {Bayesian Methods in Cosmology}

\bibitem[{Kawasaki \& Sekiguchi(2010)}]{kawasaki2010probing}
Kawasaki, M. \& Sekiguchi, T. 2010, Journal of Cosmology and Astroparticle
  Physics, 2010, 013

\bibitem[{Kilbinger {et~al.}(2010)Kilbinger, Wraith, Robert, Benabed,
  Capp{\'e}, Cardoso, Fort, Prunet, \& Bouchet}]{kilbinger2010bayesian}
Kilbinger, M., Wraith, D., Robert, C., {et~al.} 2010, Monthly Notices of the
  Royal Astronomical Society, 405, 2381

\bibitem[{{Kim} {et~al.}(2012){Kim}, {Naselsky}, \& {Mandolesi}}]{kim2012}
{Kim}, J., {Naselsky}, P., \& {Mandolesi}, N. 2012, \apjl, 750, L9

\bibitem[{King {et~al.}(2013)King, Kutyniok, \& Zhuang}]{king12}
King, E.~J., Kutyniok, G., \& Zhuang, X. 2013, Journal of Mathematical Imaging
  and Vision, to appear

\bibitem[{{March} {et~al.}(2011){March}, {Trotta}, {Berkes}, {Starkman}, \&
  {Vaudrevange}}]{March}
{March}, M.~C., {Trotta}, R., {Berkes}, P., {Starkman}, G.~D., \&
  {Vaudrevange}, P.~M. 2011, \mnras, 418, 2308

\bibitem[{{Mielczarek} {et~al.}(2009){Mielczarek}, {Szydlowski}, \&
  {Tambor}}]{Mielczarek}
{Mielczarek}, J., {Szydlowski}, M., \& {Tambor}, P. 2009, ArXiv e-prints

\bibitem[{Nikolova(2007)}]{nikolova2007}
Nikolova, M. 2007, Inverse Problems and Imaging, 1, 399

\bibitem[{{Perotto} {et~al.}(2010){Perotto}, {Bobin}, {Plaszczynski}, {Starck},
  \& {Lavabre}}]{perotto10}
{Perotto}, L., {Bobin}, J., {Plaszczynski}, S., {Starck}, J., \& {Lavabre}, A.
  2010, \aap, 519, A4+

\bibitem[{{Plaszczynski} {et~al.}(2012){Plaszczynski}, {Lavabre}, {Perotto}, \&
  {Starck}}]{stephane2012}
{Plaszczynski}, S., {Lavabre}, A., {Perotto}, L., \& {Starck}, J.-L. 2012,
  \aap, 544, A27

\bibitem[{{Rassat} {et~al.}(2012){Rassat}, {Starck}, \&
  {Dupe}}]{starck:rassat2012}
{Rassat}, A., {Starck}, J.-L., \& {Dupe}, F.~X. 2012, Submitted to A\&A

\bibitem[{{Rauhut} \& {Ward}(2012)}]{rauhut10}
{Rauhut}, H. \& {Ward}, R. 2012, Journal of Approximation Theory, 164

\bibitem[{{Starck} {et~al.}(2013){Starck}, {Fadili}, \& {Rassat}}]{starck2012}
{Starck}, J.-L., {Fadili}, M.~J., \& {Rassat}, A. 2013, \aap, 550, A15

\bibitem[{Starck {et~al.}(2010)Starck, Murtagh, \& Fadili}]{starck:book10}
Starck, J.-L., Murtagh, F., \& Fadili, M. 2010, Sparse Image and Signal
  Processing (Cambridge University Press)

\bibitem[{Trotta(2008)}]{trotta2008bayes}
Trotta, R. 2008, Contemporary Physics, 49, 71

\bibitem[{Trotta(2012)}]{Trotta:2012fk}
Trotta, R. 2012, in Springer Series in Astrostatistics, Vol.~2, Astrostatistics
  and Data Mining, ed. L.~M. Sarro, L.~Eyer, W.~O'Mullane, \& J.~De~Ridder
  (Springer New York), 3--15

\bibitem[{{Watkinson} {et~al.}(2012){Watkinson}, {Liddle}, {Mukherjee}, \&
  {Parkinson}}]{Watkinson}
{Watkinson}, C., {Liddle}, A.~R., {Mukherjee}, P., \& {Parkinson}, D. 2012,
  \mnras, 424, 313

\bibitem[{{Wehus} {et~al.}(2012){Wehus}, {N{\ae}ss}, \& {Eriksen}}]{Wehus}
{Wehus}, I.~K., {N{\ae}ss}, S.~K., \& {Eriksen}, H.~K. 2012, \apjs, 199, 15

\end{thebibliography}

\end{document}